\documentclass[reprint,superscriptaddress,amsmath,amssymb,aps,prd]{revtex4-1}

\usepackage{graphicx}
\usepackage{natbib}
\usepackage{hyperref}
\usepackage{cleveref}
\usepackage[caption=false,labelformat=empty,captionskip=0pt]{subfig}
\usepackage{todonotes}
\usepackage{mathrsfs}

\usepackage[normalem]{ulem}

\hypersetup{colorlinks=true,citecolor=blue}

\newcommand{\Esch}{E_\text{crit}}
\newcommand{\Prob}{\mathbb{P}}

\newcommand{\qed}{\text{QED}}
\newcommand{\mcsp}{\text{sc}}
\newcommand{\micron}{{\mu\mathrm{m}}}
\newcommand{\Wcmsqd}{\mathrm{W}\mathrm{cm}^{-2}}
\newcommand{\avg}[1]{\left\langle #1 \right\rangle}
\newcommand{\rmd}{\mathrm{d}}
\newcommand{\Energy}{\mathcal{E}}
\newcommand{\abs}[1]{\left|{#1}\right|}
\newcommand{\eV}{\text{eV}}

\renewcommand{\vec}[1]{\mathbf{#1}}
\DeclareMathOperator{\Ai}{Ai}

\begin{document}

\title{Benchmarking semiclassical approaches to strong-field QED:\\ nonlinear Compton scattering in intense laser pulses}

\author{T. G. Blackburn}
\email{tom.blackburn@chalmers.se}
\affiliation{Department of Physics, Chalmers University of Technology, SE-41296 Gothenburg, Sweden}
\author{D. Seipt}
\affiliation{Physics Department, Lancaster University, Lancaster LA1 4YB, UK} 
\affiliation{The Cockcroft Institute, Daresbury Laboratory, Warrington WA4 4AD, UK}
\author{S. S. Bulanov}
\affiliation{Lawrence Berkeley National Laboratory, Berkeley, California 94720, USA}
\author{M. Marklund}
\affiliation{Department of Physics, Chalmers University of Technology, SE-41296 Gothenburg, Sweden}

\date{\today}

\begin{abstract}
The recoil associated with photon emission is key to the dynamics of ultrarelativistic electrons in strong electromagnetic fields, as are found in high-intensity laser-matter interactions and astrophysical environments such as neutron star magnetospheres.
When the energy of the photon becomes comparable to that of the electron, it is necessary to use quantum electrodynamics (QED) to describe the dynamics accurately.
However, computing the appropriate scattering matrix element from strong-field QED is not generally possible due to multiparticle effects and the complex structure of the electromagnetic fields.
Therefore these interactions are treated semiclassically, coupling probabilistic emission events to classical electrodynamics using rates calculated in the locally constant field approximation.
Here we provide comprehensive benchmarking of this approach against the exact QED calculation for nonlinear Compton scattering of electrons in an intense laser pulse.
We find agreement at the percentage level between the photon spectra, as well as between the models' predictions of absorption from the background field, for normalized amplitudes $a_0 > 5$.
We discuss possible routes towards improved numerical methods and the implications of our results for the study of QED cascades.
\end{abstract}

\maketitle

\section{Introduction}

Petawatt and multipetawatt laser facilities that reach focussed intensities
in excess of ${\sim}10^{22}~\Wcmsqd$~\cite{Hercules,Pirozhkov:17}
hold great promise for the study of the interaction of charged particles
with electromagnetic fields of unprecedented strength~\cite{MourouReview,%
MarklundReview,SVBulanovNima,DiPiazzaReview}.
In these environments, the recoil associated with emission of radiation,
known as radiation reaction, can become so large that it dominates the
particle dynamics~\cite{Bulanov2004,Hadad2010}. Furthermore, when the
energy of individual photons of this radiation becomes comparable to
that of the emitting particle, it becomes essential to incorporate
quantum effects on this radiation reaction~\cite{DiPiazza2010} in our
modelling of plasmas as sources of high-energy photons~\cite{Gonoskov2017},
electron-positron pairs~\cite{BellKirk,Ridgers2012} or as laboratory
analogues of high-field astrophysical environments~\cite{Harding,Ruffini}.

However, it is not currently possible to use the most general and accurate
approach, the theory of strong-field quantum electrodynamics (QED), to model
many scenarios of interest, for reasons we will shortly outline. Instead,
a semiclassical approach has been widely adopted for use in numerical
simulations of laser-plasma and laser-particle-beam interactions.
Inherent to this model are a number of assumptions, making it essential
that we benchmark its predictions against those from QED. In this work
we do so for photon emission in the collision of ultrarelativistic
electrons with intense laser pulses. We focus on the classically
nonlinear, moderately quantum regime, motivated not only by progress
in the development of the next generation of high-intensity laser
facilities~\cite{EliWhiteBook,Apollon}, but also by recent experimental
work on radiation reaction in strong fields~\cite{Cole2018,Wistisen2018,Poder}.

Strong-field QED is not used directly to model these kinds of interactions
for a number of reasons. First, a scattering-matrix calculation connects
asymptotic free states, thereby requiring \emph{ab initio} complete
knowledge of the spatiotemporal structure of the background electromagnetic
field; exact analytical calculations have thus far proven possible only for
certain field configurations that possess high symmetry~\cite{Heinzl2017},
such as plane electromagnetic waves~\cite{Ritus} and static magnetic
fields~\cite{Erber}. Furthermore, it is generally assumed that back-reaction
effects may be neglected. This is especially important when considering
QED cascades \cite{BellKirk,Fedotov,Bulanov2010}, in which the initial state contains a single electron,
positron or photon and the final state many of these, because we expect
significant absorption of energy from the background~\cite{Seipt2017}.
(See \cite{Ilderton2018} for analysis beyond this approximation.)
Even in the absence of significant depletion, the multiplicity alone makes
calculation of a cascade within strong-field QED extremely challenging.
State-of-the-art results are those in which the final state contains
two additional particles, such as double Compton scattering~\cite{Seipt2012,%
Mackenroth2013,King2015} and trident pair creation~\cite{Hu2010,%
Ilderton2011,King2013,Dinu2018} from single electrons.

These conditions, namely complex field structure, strong depletion due to
back-reaction and high multiplicity, are ubiquitous in the interaction
of high-intensity lasers with particle beams or plasma targets. As such,
considerable effort has been devoted to the development of numerical
schemes that can model QED cascades as well as self-consistent plasma
dynamics~\cite{Duclous,Elkina}. We characterize these as \emph{semiclassical},
because they factorize the cascade into a product of first-order processes
that occur in vanishingly small regions linked by classically determined
trajectories; the production rates and spectra are calculated for the
equivalent QED process in constant, crossed fields~\cite{Ridgers2014}.
This is possible because at high intensity (to be defined in \cref{sec:Methods}),
the formation length over which a photon is emitted, or an electron-positron
pair is created, is much smaller than the length scale over which the
field varies~\cite{Ritus}.
The approximation that emission occurs instantaneously, and therefore
that the rate for a constant, crossed field may be employed, is called
the \emph{locally constant field approximation} (LCFA). Monte Carlo
implementations of QED processes based on this have found widespread
adoption in particle-in-cell (PIC) codes (see \cite{Gonoskov2015} for
details). Depletion in these codes is therefore treated classically, as
QED processes alter the plasma current density $\vec{j}$, which alters
the energy density of the self-consistent electric and magnetic fields
$\vec{E}$ and $\vec{B}$ via the $\vec{j} \cdot \vec{E}$ term in Poynting's
theorem; see \cite{Nerush,Grismayer,DelSorbo} for examples of how this
drives laser absorption.

Identifying the parameter regime where this semiclassical picture works,
and why, has been the subject of much theoretical work~\cite{Dinu2016,%
Meuren,DiPiazza2017}. However, there has been limited direct benchmarking
of numerical and analytical results in regimes of experimental interest.
For nonlinear Compton scattering (photon emission by an electron),
\citet{Harvey2015} compared the frequency and angular spectra predicted
by 1) integration of the QED probability rate for a monochromatic plane
wave and 2) semiclassical simulation of a 100~fs pulsed plane wave with
super-Gaussian temporal profile, concluding that the neglect of
interference effects in the latter caused harmonic structure to be missed.

In this work we present systematic comparisons not only of the
longitudinal and transverse momentum spectra (\cref{sec:Lightfront,%
sec:Perpendicular}), but also the absorption of energy from the background
field (\cref{sec:Absorption}).
We introduce a normalization framework in \cref{sec:ComparisonBasis}
that guarantees that we compare precisely the same physical scenario.
This permits direct, quantitative benchmarking of semiclassical codes
against analytical results from QED in the parameter regime relevant
for recent and upcoming experiments.

\section{Methods}
\label{sec:Methods}

	\begin{figure}
	\includegraphics[width=0.66\linewidth]{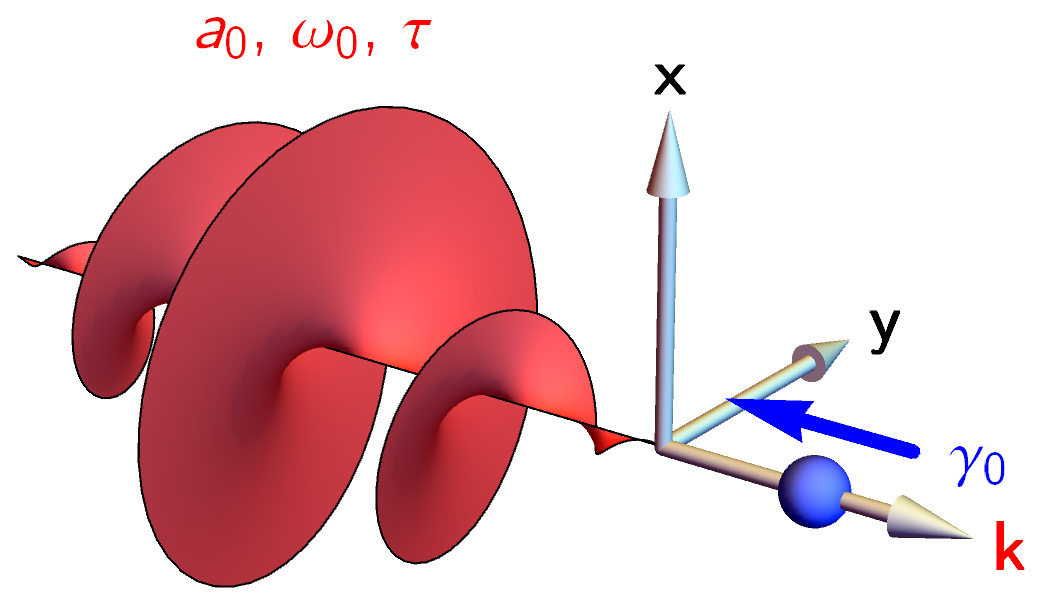}
	\caption{%
			An electron (blue) with initial Lorentz factor $\gamma_0 \gg 1$
			collides head-on with an intense, circularly polarized laser
			pulse (red) that has strength parameter $a_0$, angular
			frequency $\omega_0$ and duration $\tau$.}
	\label{fig:Diagram}
	\end{figure}

The interaction geometry is illustrated in \cref{fig:Diagram}.
An electron with initial Lorentz factor $\gamma_0$ collides head-on
with a circularly polarized laser pulse that has dimensionless
amplitude $a_0$, central frequency $\omega_0$ and
invariant duration $\tau$. Throughout this work we set $\hbar = c = 1$
and denote the elementary charge by $e$ and the electron mass by $m$.
The pulse vector potential $e A^\mu(\phi) = m a_0 g(\phi) (0, \sin\phi, \cos\phi, 0)$,
where $a_0 = e E_0 / (m \omega_0)$ for electric field strength $E_0$~\cite{Heinzl2009}
and $g(\phi) = \cos^2[\phi/(4\tau)]$ for phases $\abs{\phi} < 2 \pi \tau$.
In all results presented here, $\gamma_0 = 1000$ and $\omega_0 = 1.55\,\eV$
(equivalent to a wavelength $\lambda = 0.8~\micron$). We will consider
dimensionless amplitudes in the range $5 \leq a_0 \leq 30$, which covers the
transition between the weakly and highly nonlinear classical regimes,
and restrict the laser duration to be $\tau = 2$ or 3 so that the
expected number of photons is of order one. This is because our QED
calculations are performed for single scattering only, and so that
we can gather sufficient statistics in the semiclassical simulations
(the fraction of collisions in which only one photon is emitted is
exponentially suppressed with increasing $a_0$).

The quantum interaction of charged particles and photons with strong
fields is characterized by the invariants $\chi_e = e\sqrt{-(F.p)^2}/m^3$
and $\chi_\gamma = e\sqrt{-(F.k')^2}/m^3$, where $F$ is the electromagnetic
field tensor and $p$ and $k'$ the four-momenta of the electron and photon
respectively~\cite{Ritus}. $\chi_e$ may be interpreted as a measure of 
the field strength in the rest frame of the electron relative to that
of the critical field of QED $\Esch = m^2/e$~\cite{Sauter,Heisenberg,Schwinger}.
It is often referred to as the `quantum nonlinearity parameter' by
analogy with $a_0$, which is the classical nonlinearity parameter~\cite{Ritus}. We have
$\chi_e \sim 0.1$ for the interaction parameters under consideration
here so quantum effects are non-negligible.

\subsection{QED}
\label{sec:qed}

\begin{figure}
	\centering
	\includegraphics[width=\linewidth]{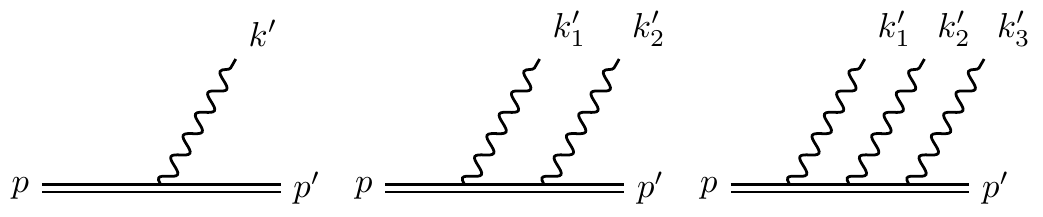}
	\caption{%
		Feynman diagrams for the emission of photons with four-momenta
		$k'_{1,2,\ldots}$ in the scattering of a laser-dressed electron
		from asymptotic four-momentum $p$ to $p'$. The double lines
		indicate that the interaction with the background field is
		calculated to all orders in the coupling $a_0$.}
	\label{fig:Feynman}
	\end{figure}

The strong-field QED scattering matrix (S-matrix) connects asymptotic free
states, evolving the initial state from the distant past to the distant future.
The calculation is performed to all orders in the coupling to the background
field $a_0$, i.e. non-perturbatively, as the number of photons
absorbed and reemitted by an electron in an intense laser field is very large.
For the lowest order process shown in \cref{fig:Feynman}, the emission of
one photon or single nonlinear Compton scattering
\cite{Narozhnyi:JETP1965,Boca:PRA2009,Krajewska:PRA2012,Seipt2016}, it reads
	\begin{equation}
	S_1 =
		-i e (2\pi)^3
		\delta_\mathrm{lf} (p' + k' - p)
		\sum\nolimits_j \! \mathscr{T}_j \mathscr{C}_j.
	\end{equation}
The delta function ensures the conservation of momentum in the lightfront
and transverse directions. By \emph{lightfront} momentum we mean
$p^+ \equiv k.p/\omega_0$, which is conserved in a plane wave with
wavevector $k$ in the absence of radiation reaction.
Other features are the transition operators
$\mathscr{T}_j$ which are sensitive to the electron spins and the
photon polarization, and the $\mathscr{C}_j$ which are integrals over
the laser phase (see \citet{Seipt2016} for details)
	\begin{equation}
	\mathscr{C}_j =
		\int \!
		\rmd \phi \,
			\mathscr{F}_j(\phi)
			e^{i \int \! \rmd\phi \, \frac{k'. \pi (\phi )}{k.p'}}.
	\end{equation}
Here the $\mathscr{F}_j$ are functions of the vector potential $A(\phi)$ and
$\pi(\phi)$ in the exponent is the classical kinematic four-momentum of
the electron, a solution to the radiation-free Lorentz force equation.

The one-photon emission probability is
	\begin{equation}
	\frac{\rmd^3 \Prob_1}{\rmd f \rmd^2 \vec{r}_\perp} =
		\frac{\alpha m^2}{(4 \pi \omega_0 p^+)^2}
		\frac{f}{1-f}
		\,
		\frac{1}{2}
		\sum_\mathrm{spin,pol}
		\abs{	 \sum\nolimits_j \! \mathscr{T}_j \mathscr{C}_j }^2
	\end{equation}
where we have defined the lightfront momentum transfer fraction
$f = k'^+ /p^+$ and normalized transverse photon momentum
$\vec{r}_\perp = \vec{k}'_\perp / (f m)$. The magnitude of the latter
$r_\perp = (p^+/m) \tan(\theta/2)$, where $\theta$ is the polar angle
of the emitted photon, becoming $r_\perp \simeq \gamma \theta$ if
$\gamma \gg 1$ and $\theta \ll 1$.

If $a_0^3/\chi_e \gg 1$ and $f$ is not too small, the phase interval
which contributes to the emission of a single photon becomes much smaller
than the wavelength of the background field~\cite{Ritus} and
interference between emission from different formation regions is
suppressed~\cite{Dinu2016}. In this case, the field may be treated
as constant over the photon formation region.
As the photons are emitted into a narrow
cone around the direction of the electron's instantaneous momentum,
we can integrate over the transverse momenta $\vec{r}_\perp$ to
obtain the instantaneous probability rate per unit phase and lightfront
momentum transfer
	\begin{equation}
	\frac{\rmd W}{\rmd f} =
		-\frac{\alpha m^2}{\omega_0 p^+}
		\left[
			\Ai_1(z) +
			\left(\frac{2}{z} +  \chi_e f \sqrt{z} \right) \Ai'(z)
		\right],
	\label{eq:LCFARate}
	\end{equation}
where $z^{3/2} = f/[\chi_e (1-f)]$ and $\chi_e \equiv \chi_e(\phi)$ the
local value of the quantum parameter.

\subsection{Semiclassical}
\label{sec:Semiclassical}

In the semiclassical interpretation of the collision process, the electron
follows a (radiation-free) classical trajectory between point-like,
probabilistically determined, QED events. These events are implemented
using the standard Monte Carlo algorithm~\cite{Ridgers2014,Gonoskov2015},
with rates calculated in the LCFA, i.e. \cref{eq:LCFARate} (see also \cite{Ritus,Erber,BKS}).
We use \texttt{circe}, a particle-tracking code that simulates photon and
positron production by high-energy electrons (and photons) that collide
with laser pulses that have $a_0 \gg 1$.
In one spatial dimension the external field is assumed to be a plane
electromagnetic wave, i.e. the fields are determined by a single parameter
$\phi = \omega_0 n.x$ where $\omega_0$ is the wave frequency,
$n = (1,\hat{\mathbf{k}})$ for direction of propagation $\hat{\mathbf{k}}$,
and $x$ is the four-position of the electron. Between emissions, the
electron dynamics are determined by the Lorentz force alone. The spatial
components of the four-momentum $p = (\gamma m, \mathbf{p})$ that are
perpendicular to the wavevector are obtained by integrating
	\begin{equation}
	\frac{\rmd \vec{p}_\perp}{\rmd \phi} =
		-\frac{e \vec{E}_\perp (\phi)}{\omega_0}
	\end{equation}
where $\vec{E}_\perp$ is the electric field at phase $\phi$. The
remaining components of $p$ are determined by the two conditions
$p^+ = \text{const}$ and $p^2 = m^2$. The four-position is determined by
$\frac{\rmd x}{\rmd \phi} = p/(\omega_0 p^+)$.

Photon emission is implemented as follows. Each electron is assigned an
optical depth against emission $\tau = -\log(1-R)$ for pseudorandom
$0 \leq R < 1$, which evolves as
	\begin{equation}
	\frac{\rmd \tau}{\rmd \phi} =
		-{\int_0^1\! \frac{\rmd W}{\rmd f} \,\rmd f},
	\end{equation}
where $W$ is the instantaneous probability rate of emission given by
\cref{eq:LCFARate}, until the point where it falls below zero.
Then the lightfront momentum transfer $f = \chi_\gamma / \chi_e$ is
pseudorandomly sampled from the differential rate and $\tau$ is reset.
Assuming that emission occurs in the direction parallel to the initial
momentum, as the electron emits into a narrow cone of opening angle
$1/\gamma$, the momenta of the electron and photon after the scattering
are
	\begin{align} \label{eq:MomentumAssignment}
	\begin{split}
	p' &= (m \sqrt{1 + (1-f)^2 (\gamma^2-1)}, (1 - f) \mathbf{p}),
	\\
	k' &= (f m \sqrt{\gamma^2 - 1}, f \mathbf{p}). 
	\end{split}
	\end{align}
As discussed by \cite{Duclous}, this leads to an error
in energy conservation of
	\begin{equation}
	\Delta \Energy = \frac{m}{2\gamma} \frac{f}{1-f} + O(\gamma^{-3})
	\label{eq:EnergyError}
	\end{equation}
which is small for ultrarelativistic particles.


\subsection{Comparison basis}
\label{sec:ComparisonBasis}

In this work we present quantitative, as well as qualitative, comparisons
of electron and photon spectra predicted by the exact QED and semiclassical
methods. We discuss here how the normalization may be set consistently,
but independently by the two methods.

The final result of the QED calculation is the probability $\Prob_1$ that
a single photon is emitted in collision of an electron with a pulsed
electromagnetic plane wave. However, even for the short pulses under
consideration here, the fact that $a_0 > 1$ makes it possible for
$\Prob_1 > 1$. Where this occurs the probability is generally interpreted
as the mean number of emitted photons, i.e.
$\Prob_1 \rightarrow N_{\gamma,\qed}$, as this quantity can certainly
exceed unity~\cite{Peskin,DiPiazza2010,Harvey2015}. We emphasise that this
interpretation is exact only in the classical limit, where recoil can be
neglected. The true probability for single scattering is given by the
renormalized quantity $\Prob_1 / (1 + \sum^\infty_{n=1} \Prob_n)$.
To determine this would require the calculation of the scattering
probability to a state containing an arbitrary number of photons $n$.
Efforts to characterize such multiphoton interactions analytically have
been limited due to the complexity of the calculations; at present all
results in the literature are restricted to $n \leq 2$.
For these reasons, we will define the QED `number of photons' as
	\begin{equation}
	N_{\gamma,\qed} \equiv
		\int \!
			\frac{\rmd^3 \Prob_1}{\rmd f \rmd^2 \vec{r}_\perp}
		\, \rmd f \rmd^2 \vec{r}_\perp.
	\label{eq:NumberQED}
	\end{equation}

In the semiclassical calculation, multiphoton emission is accounted for
by factorisation of the multiphoton emission into a product of first-order processes.
Localizing emission in this way allows us to determine the branching
fraction to a final state containing an arbitrary number of particles,
thereby guaranteeing that $\Prob_1 < 1$. In the classical limit (i.e.
negligible recoil per photon), one emission event is independent of any other and
the probability that $n$ photons are emitted in a given collision
$\Prob_n = \lambda^n e^{-\lambda}/n!$, where
$\lambda \equiv N_{\gamma,\mcsp}$, the mean number of photons
in the semiclassical case. However, we consider here collisions where
$\chi \sim 0.1$ and recoil is not negligible.
As the emission rate (at fixed field strength) decreases with increasing
particle energy, emitting a photon and so losing energy makes it more
probable that further photons are emitted. As such, the numbers of photons
emitted in two non-overlapping intervals are not independent and the
probability $\Prob_n$ ceases to be Poisson-distributed.
In summary, radiation reaction, the recoil due to photon
emission, affects the average number of photons because ``the emission of
each photon modifies the electron state and, consequently, the next
emissions''~\cite{DiPiazza2010}.

Since it is not possible, as yet, to determine the renormalization factor
by which the QED results should be scaled, we propose this alternative.
The QED results from \cref{eq:NumberQED} are \emph{not} scaled.
Equivalent semiclassical spectra are obtained statistically, by
generating a large set of collision data, accepting only those collisions
in which exactly one photon is emitted, and rescaling such
that the spectra have integral $N_{\gamma,\mcsp}$. The mean number
of photons $N_{\gamma,\mcsp}$ is determined by considering the entire set of
collision data, i.e.
	\begin{equation}
	N_{\gamma,\mcsp} \equiv
		\frac{\sum_i i N_i}{\sum N_i}
		\frac{1}{N_1}
		\int \! \frac{\rmd^2 N_1}{\rmd f \rmd r_\perp} \, \rmd f \rmd r_\perp,
	\label{eq:NumberMCSP}
	\end{equation}
where $N_i$ is the number of simulated collisions in which exactly $i$ photons
are emitted.

This definition ensures that only collisions with a single
emission contribute to the shape of the spectrum, and that its integral
may be interpreted in a manner consistent with the QED result. From now on,
all differential spectra will be given in terms of the `number of photons'
defined by \cref{eq:NumberQED,eq:NumberMCSP}.
We note that while this post-facto selection criterion lets us compare
the same physical scenario as the QED approach, multiphoton and recoil
corrections are still present because $N_{\gamma,\mcsp}$ is
affected by radiation reaction. We do not necessarily expect the QED
`probability' $\Prob_1$ to satisfy
$\Prob_1 \rightarrow N_{\gamma,\qed} = N_{\gamma,\mcsp}$
for this reason.

\section{Results}

\subsection{Lightfront momentum}
\label{sec:Lightfront}

	\begin{figure*}
	\subfloat[]{\label{fig:NP-a}\includegraphics[width=0.66\linewidth]{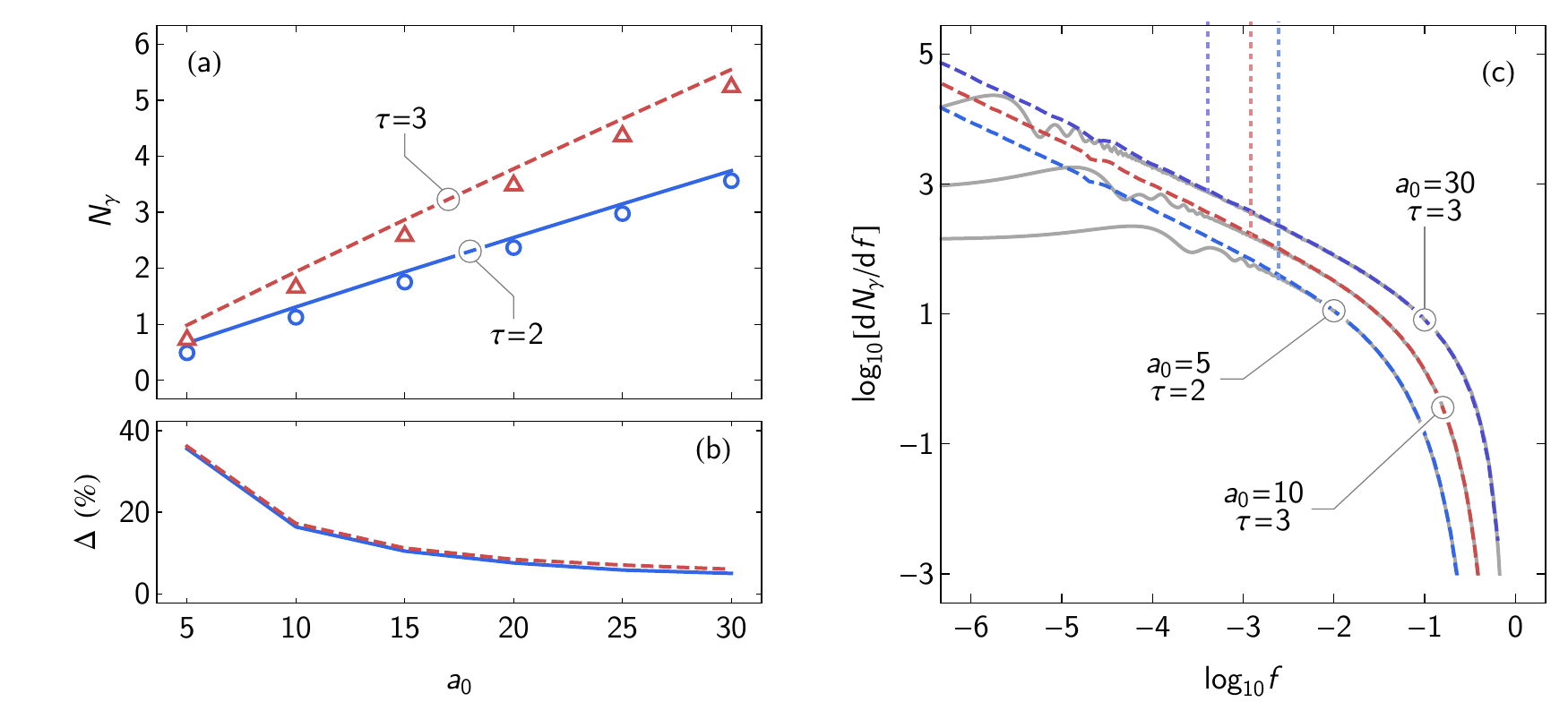}}
	\subfloat[]{\label{fig:NP-b}}
	\subfloat[]{\label{fig:NP-c}}
	\caption[Number of photons]
		{(a) The number of photons emitted in the collision of
		an electron with $\gamma_0 = 1000$ and a laser pulse with given $a_0$
		and duration $\tau$, calculated analytically from QED (points) and
		from semiclassical simulation (lines).
		(b) The percentage difference between the number of photons
		as evaluated by the two methods.
		(c) The lightfront-momentum spectrum for the specified collision
		parameters: results from QED (solid, grey) and semiclassical simulation
		(dashed, coloured). Vertical, dotted lines indicate $f_C$,
		the first Compton edge of a monochromatic plane wave with
		the same $a_0$.}
	\label{fig:NumberOfPhotons}
	\end{figure*}

	\begin{figure*}
	\subfloat[]{\label{fig:LF-a}\includegraphics[width=\linewidth]{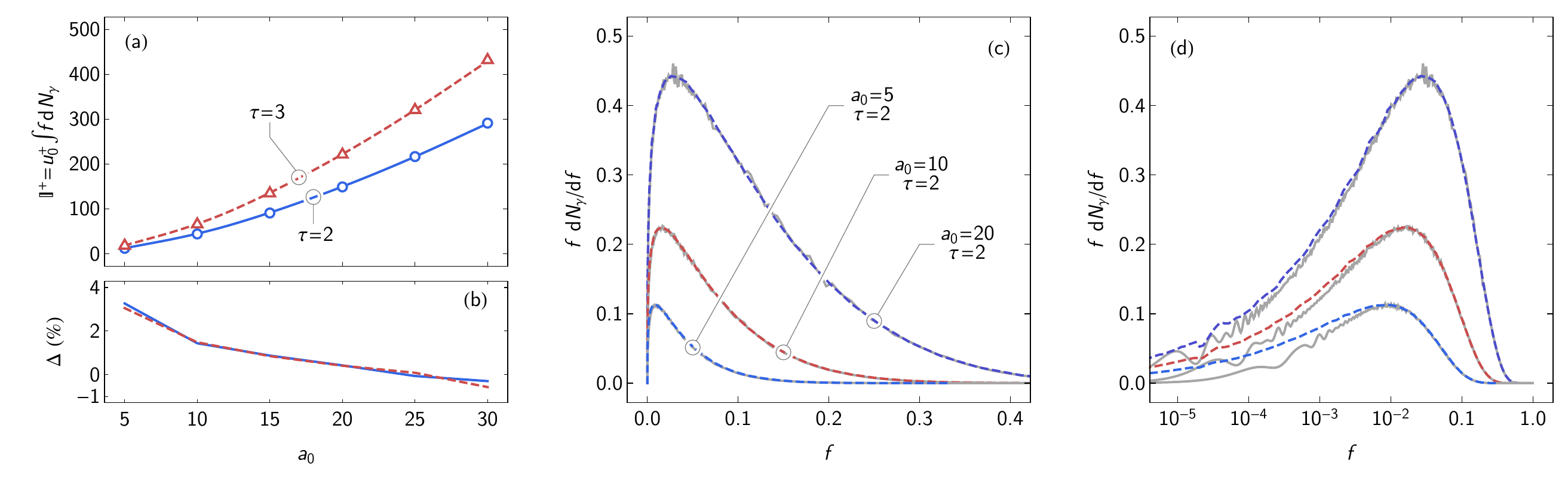}}
	\subfloat[]{\label{fig:LF-b}}
	\subfloat[]{\label{fig:LF-c}}
	\subfloat[]{\label{fig:LF-d}}
	\caption[Lightfront momentum]
		{(a) The total lightfront momentum lost by an electron with
		$\gamma_0 = 1000$ 	in a collision with a laser pulse with given
		$a_0$ and duration $\tau$, calculated analytically from QED (points)
		and from semiclassical simulation (lines).
		(b) The percentage difference between the two.
		(c, d) The lightfront-momentum intensity spectrum for the specified
		collision parameters: results from QED (solid, grey) and
		semiclassical simulation (dashed, coloured).}
	\label{fig:Lightfront}
	\end{figure*}

The symmetries of a plane electromagnetic wave make lightfront momentum
$u^+ \equiv n.u$ a natural choice of parametrization for the differential
scattering probability, because the conservation of momentum for (single)
nonlinear Compton scattering reads $u^+_0 = u^+_e + u^+_\gamma$, where
$u^+_e = p'^+/m$ and $u^+_\gamma=k'^+/m$ are the normalized lightfront momenta of the scattered electron
and photon respectively. We will plot differential spectra in terms of the
transfer fraction $f = u^+_\gamma / u^+_0 = 1 - u^+_e / u^+_0$. In the
back-scattering limit, $f \simeq \omega'/(\gamma_0 m)$ where $\omega'$ is
the energy of the scattered photon.

We compare the analytical and simulation predictions for the number of
photons emitted in the head-on collision of an electron with initial energy
$\gamma_0 = 1000$ and a two- or three-cycle laser pulse in \cref{fig:NP-a},
with examples of the differential spectra in \cref{fig:NP-c}. The percentage
difference between the results of the two methods is given for the
total number of photons in \cref{fig:NP-b}.
We find that the semiclassical method systematically overestimates the
number of photons, but that the fractional discrepancy diminishes with
increasing $a_0$, falling below 10\% when $a_0 \geq 20$. $N_\gamma$ scales
approximately linearly with $a_0$ as expected in the strong-field regime;
the growing discrepancy towards the lower end of the plotted range is an
indication of the transition to the perturbative regime where
$N_\gamma \propto a_0^2$ instead.

The origin of the discrepancy is shown in \cref{fig:NP-c}. While there
is very good agreement for large $f$, i.e. high energy, the semiclassical
method strongly overestimates the number of photons with vanishing $f$.
This is because the underlying LCFA rate contains an integrable singularity
$\propto f^{-2/3}$ absent in the exact QED calculation~\cite{Dinu2012}.
In the latter case, the probability tends to a finite value~\cite{DiPiazza2017}
	\begin{equation}
	\lim_{f \rightarrow 0} \frac{\rmd \Prob_1}{\rmd f} =
		\frac{\alpha a_0^3}{2 \chi_e} \int \!\rmd \phi \,  g^2 (\phi)
	\end{equation}
where $g(\phi)$ is the envelope function given in \cref{sec:Methods}.
It is not surprising that the semiclassical spectra do not reproduce this
limit, because the LCFA arises from an expansion in the parameter
$\chi_e/a_0^3 \ll 1$~\cite{Dinu2016}.

The physical meaning of this parameter is that it compares the formation
length of the photon to the length scale over which the field varies.
If this is sufficiently small, we can assume emission occurs instantaneously
and thereby neglect interference effects. As discussed in \citet{Harvey2015},
this means that Monte Carlo implementations of localized rates cannot
reproduce the harmonic structure that becomes visible in the emission
spectrum at small $f$. A simple way to estimate the smallest $f$ for
which the LCFA should be valid is to recall that in a monochromatic plane
wave, emission over a complete phase oscillation contributes to photons
at the first nonlinear Compton edge, for which the transfer fraction
$f_C \simeq 2 \chi_e / a_0^3$.
The requirement that the formation length be smaller than the laser
wavelength is then equivalent to having $f \gtrsim f_C$. This limit is
consistent with the results shown in \cref{fig:NP-c} and with a more
detailed calculation performed by \citet{DiPiazza2017}. In fact, if we
cut off the QED and semiclassical spectra below $f = f_C$, the percentage
discrepancy in the number of photons falls from 17\% to 5\% at $a_0 = 10$
and from 5\% to 2\% at $a_0 = 30$.

	\begin{figure*}
	\subfloat[]{\label{fig:PM-a}\includegraphics[width=\linewidth]{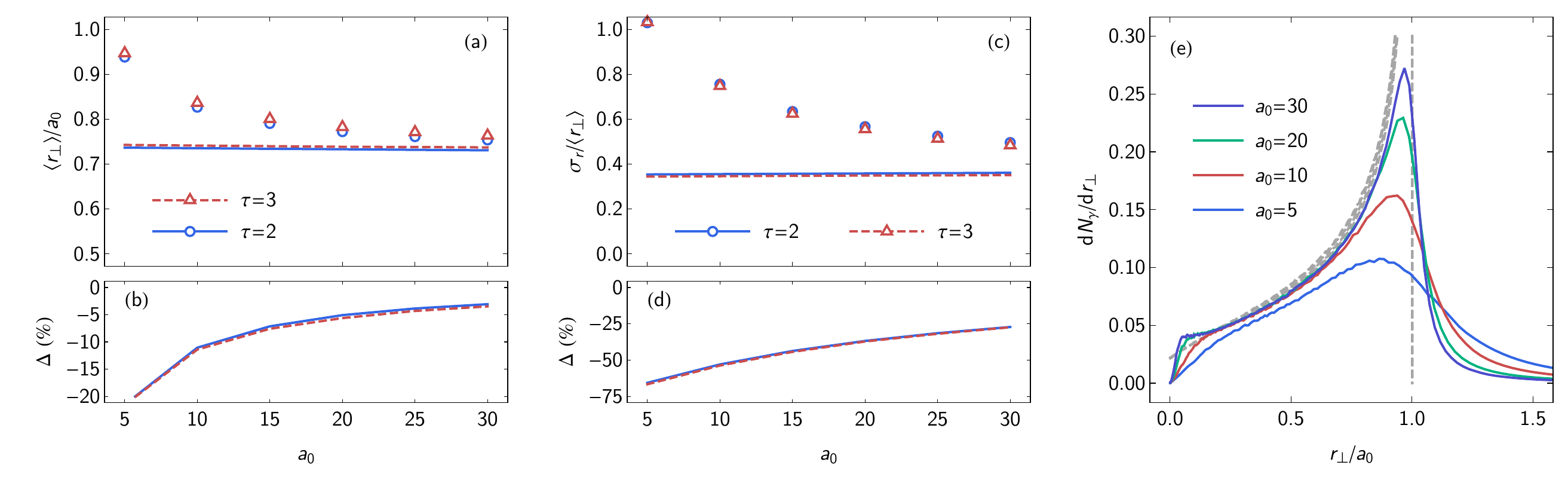}}
	\subfloat[]{\label{fig:PM-b}}
	\subfloat[]{\label{fig:PM-c}}
	\subfloat[]{\label{fig:PM-d}}
	\subfloat[]{\label{fig:PM-e}}
	\caption[Perpendicular momentum]
		{(a) The mean $r_\perp$ of the photon emitted in the collision of
		an electron with $\gamma_0 = 1000$ and a laser pulse with given
		$a_0$ and duration $\tau$, calculated analytically from QED (points)
		and from semiclassical simulation (lines).
		(b) The percentage difference between the two.
		(c) The standard deviation of $r_\perp$, normalized to the mean,
		and (d) the percentage difference between the QED and semiclassical
		results.
		(e) Differential $r_\perp$ spectra for the specified collision
		parameters: results from QED (solid, coloured) and semiclassical
		simulation (dashed, grey).}
	\label{fig:Perpendicular}
	\end{figure*}

While it is important to capture the number spectrum accurately, the
dynamically significant quantity is the spectrum weighted by $u^+_\gamma$,
as this gives the momentum change of the electron, or radiation
reaction~\cite{Ridgers2017,Niel2017}.
When we compare the total radiated lightfront momentum
	\begin{equation}
	\mathbb{I}^+ \equiv
		u^+_0 \int\! f \frac{\rmd N_\gamma}{\rmd f} \, \rmd f
	\label{eq:TotalLightfront}
	\end{equation}
in \cref{fig:LF-a} we find much better agreement between the two methods,
with a relative discrepancy below 4\%
even for $a_0 = 5$. This is because the contribution to the electron recoil
from the low-energy tail, i.e. the part of the spectrum where the LCFA fails,
is very small and diminishes with increasing $a_0$.
This can also be seen in \cref{fig:LF-c,fig:LF-d} where we show the
intensity spectra without log scaling on the vertical axis.

\subsection{Perpendicular momentum}
\label{sec:Perpendicular}
	
We parametrize the perpendicular momentum spectrum using the scaled
quantity $r_\perp \equiv u_\gamma^\perp / f$. For $\gamma_0 \gg a_0$ and
$\gamma_0 \gg 1$ as we have here, $r_\perp \simeq \gamma_0 \theta$ where
$\theta$ is the photon scattering angle. The comparison between the
analytical and simulation results, shown in \cref{fig:Perpendicular},
is for $r_\perp$ scaled by the laser strength parameter $a_0$ for the
following reason.

Analysis of nonlinear Compton scattering in a monochromatic, circularly
polarized plane wave only in terms of the number of laser photons absorbed
has been shown to reproduce the classical result that the photons are
typically emitted along the direction of the instantaneous momentum of
the electron in the electromagnetic field~\cite{Seipt2017}. Assuming the electron and
the laser were initially counterpropagating and that both the electron
Lorentz factor $\gamma_0$ and the laser amplitude $a_0 \gg 1$, we have
that the most probable angle of emission is
$\tan\theta = 4 a_0 \gamma_0 / (4 \gamma_0^2 - a_0^2) \simeq a_0/\gamma_0$.

In the semiclassical calculation, we capture the electron's transverse
oscillation directly by solving the classical equations of motion and
rely on relativistic beaming to justify setting the photon's emission
direction to be parallel to the electron's instantaneous momentum.
We may derive a scaling relation for the average $r_\perp$ for the pulsed
plane waves under consideration here within the framework of the LCFA.
The instantaneous angle between the electron momentum and the laser axis
is $\theta(\phi) \simeq a_0 g(\phi)/\gamma_0$ for $\gamma_0 \gg a_0 \gg 1$,
where $g(\phi)$ is the pulse envelope described in \cref{sec:Methods}.
Assuming that the photon is emitted parallel to the electron momentum,
we have that the mean value of $r_\perp$
	\begin{equation}
	\avg{r_\perp} =
		\frac{\gamma_0 \int\! \theta(\phi) W(\phi) \,\rmd\phi}{\int\! W(\phi) \,\rmd\phi}
		\simeq \frac{3 a_0}{4}
	\end{equation}
where $W(\phi)$ is the emission rate [\cref{eq:LCFARate}] integrated
over all $f$, and we obtain an analytical result by working in the classical
limit $\chi_e \ll 1$.
In this expression the mean $r_\perp$ is normalized to the number of photons.
Consequently if $\chi$ is not too large, we expect the $\avg{r_\perp}/a_0$
predicted by semiclassical simulation to be independent of $a_0$ and the
pulse duration $\tau$. This is indeed what is shown in \cref{fig:PM-a}.
The exact QED results are generally larger, but tend towards the
semiclassical results as $a_0$ is increased. This is because photons
are emitted into a broader range of angles in the QED calculation,
which can be seen by the fact that the standard deviations of the
spectra (normalized to the mean) shown in \cref{fig:PM-c}, i.e the
widths of the distribution, are larger in the analytical case.

These integral comparisons lead us to expect important qualitative
differences between the $r_\perp$ spectra predicted by QED and by
semiclassical simulation. Indeed, \cref{fig:PM-e} shows that, unlike
the former, the latter exhibit a universal shape when plotted as a
function of $r_\perp/a_0$. Furthermore, they diverge as
$r_\perp \rightarrow a_0$. No photons are emitted with $r_\perp > a_0$.
The range of accessible $r_\perp$ is much larger for the QED spectra,
though we note that as $a_0$ increases, the peak of the spectrum
(i.e. the most probable $r_\perp$) tends towards $a_0$ and the spectra
generally become narrower. As we have scaled $r_\perp$ down by $a_0$,
this indicates that the characteristic width of the spectrum is
approximately constant for all $a_0$.

	\begin{figure*}
	\subfloat[]{\label{fig:Ab-a}\includegraphics[width=0.66\linewidth]{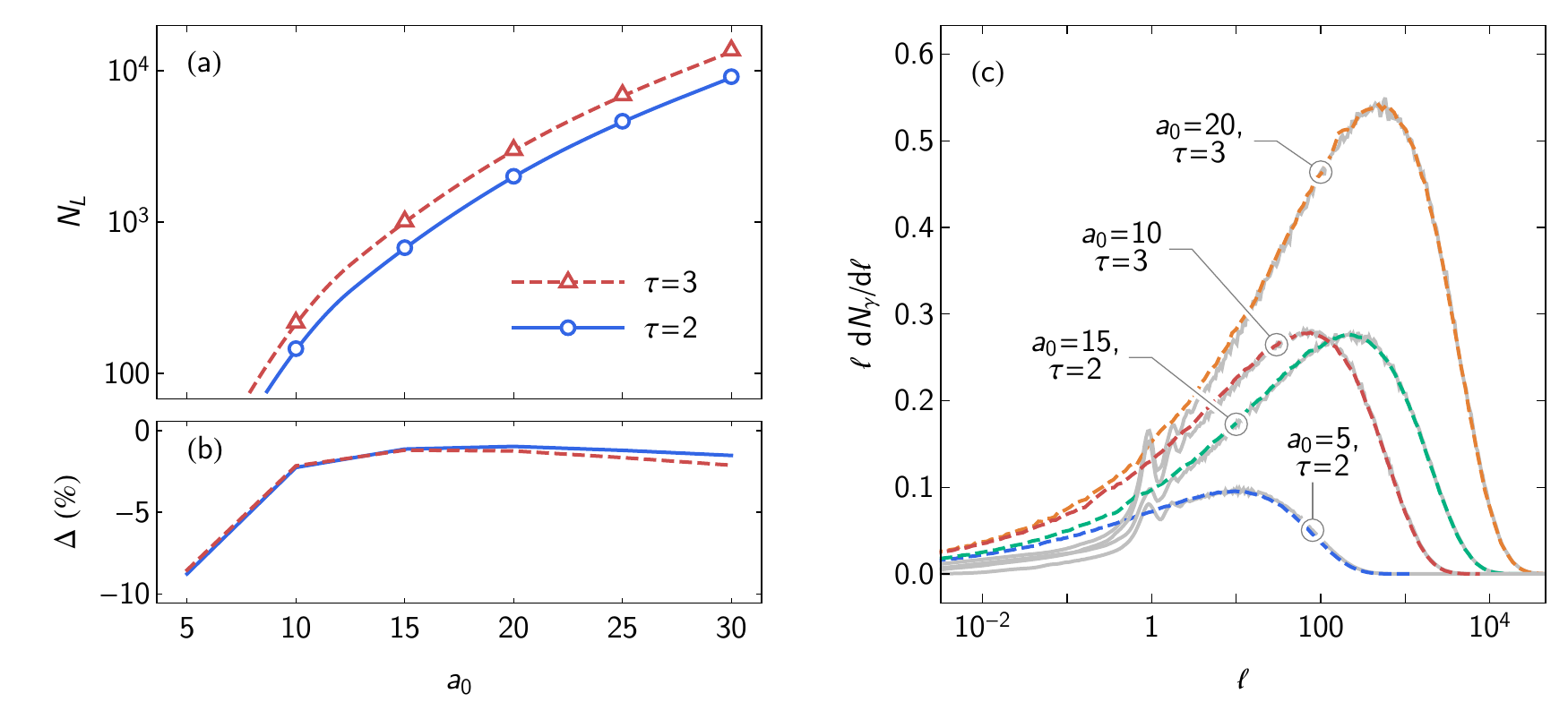}}
	\subfloat[]{\label{fig:Ab-b}}
	\subfloat[]{\label{fig:Ab-c}}
	\caption[Number of laser photons absorbed]
		{(a) The number of laser photons absorbed in the emission of a
		single $\gamma$ photon when an electron with $\gamma_0 = 1000$
		collides with a laser pulse with given $a_0$ and duration $\tau$:
		results from QED (points) and semiclassical simulation (lines).
		(b) The percentage difference between the two.
		(c) The weighted probability that $\ell$ photons are absorbed,
		from QED (grey, solid) and semiclassical simulation (coloured,
		dashed).}
	\label{fig:Absorption}
	\end{figure*}

The cause of these differences is the assumption of collinear emission
in the semiclassical simulations, i.e. assigning the final momenta
according to \cref{eq:MomentumAssignment}.
Relativistic beaming means that the
photon is emitted forward into a cone of half-angle ${\sim}1/\gamma_0$,
corresponding to a width in $r_\perp$ of $\sigma_r \sim 1$. We have
neglected this extra angular divergence, which is why the semiclassical
results have a sharp edge at $r_\perp = a_0$. The range of angles at
which a photon can be emitted is bounded by the angle between the
electron's instantaneous momentum and the laser wavevector, which is
at most $a_0/\gamma_0$ for the circularly polarized pulses under
consideration here.

A more subtle discrepancy may be seen for small $r_\perp$ in
\cref{fig:PM-e}. Whereas the analytical results tend to zero as $r_\perp$
does, exhibiting a pronounced shoulder as they do so, all the semiclassical
spectra tend to a finite value of approximately $0.022$.
(This is consistent with a classical calculation of the angular
spectrum, which gives $\rmd N_\gamma/\rmd r_\perp \simeq 5\alpha/\sqrt{3}
\simeq 0.021$ for $r_\perp = 0$.)
The difference is a consquence of the LCFA: recall that in the semiclassical
approach, photons are emitted parallel to the electron's instantaneous
momentum. Therefore photons with small $r_\perp$ originate from the
leading and trailing edges of the pulse, where the local field strength
is small and so too is the angle between the electron trajectory and
the laser wavevector. In these regions, the effective $a_0$ is small
enough that interference effects become important, suppressing photon
emission.

\subsection{Absorption}
\label{sec:Absorption}

Energy-momentum conservation demands that the emission of a photon by an
electron in a strong background field be accompanied by the absorption
of energy from that background field. As the background under consideration
here is an electromagnetic wave, this can be interpreted as the absorption
of a certain number $\ell$ of photons. \Citet{Seipt2017} have shown that
in a circularly polarized, monochromatic plane wave with strength parameter
$a_0$, the emission of a photon with quantum parameter $f = \chi_\gamma/\chi_e$
is associated with the absorption of $\ell = s$ laser photons, where
	\begin{equation}
	s = \frac{a_0^3}{\chi_e} \frac{f}{1 - f}.
	\label{eq:MostProbableNumber}
	\end{equation}
For the short pulses in this work, the probability that $\ell$ photons
are absorbed is determined numerically.

In the semiclassical method the background field is treated entirely
classically. Neverthless we may define an equivalent number of absorbed
photons by dividing the classical work done on the moving charge by
the laser frequency $\omega_0$:
	\begin{equation}
	\ell =
		-\frac{1}{\omega_0} \int \! e \mathbf{v}\cdot\mathbf{E} \, \rmd x^0.
	\label{eq:ClassicalEll}
	\end{equation}
Note that for a plane wave, the above integral would be identically zero
in the absence of radiation. (The same result holds in the QED calculation:
if no photon is emitted, $\ell = 0$.) \texttt{circe} computes
\cref{eq:ClassicalEll} for each test electron, integrating the work
done across the entire trajectory of the electron in the pulse.

A comparison between the total number of absorbed photons
$N_L \equiv \int\! \ell \,\rmd N_\gamma$, as computed by the two methods,
is shown in \cref{fig:Ab-a}. We find that the semiclassical method
systematically underestimates the absorption, but that this difference
occurs at the level of a few percent and decreases with increasing $a_0$.
This is in contrast to what we found for the number of photons and the
total radiated lightfront momentum, where the semiclassical result was
generally larger than the QED result. In all three cases we expect
errors to arise due to the finite size of the formation length and
the associated interference; however, here, our results imply that
there is some `missing' absorption.

In their analysis of electron-positron pair creation by a photon in a
strong laser field, \citet{Meuren} divided the absorbed energy into
`classical' and `quantum' parts, the former being the work done
accelerating the daughter particles out of the laser pulse
and the latter the absorption of photons over the formation length.
They showed that the classical part scales approximately like
$a_0^3/\chi_e$ and the quantum part like $a_0/\chi_e$, concluding that
the classical absorption should be dominant at high intensity. This is
consistent with the results presented here, in that we capture
the acceleration of the electron post-emission but not any absorption
over the formation length, which is assumed to be vanishingly small.
Recall that in the semiclassical simulations there is an error
in energy conservation due to the assumption of collinear emission
(see \cref{sec:Semiclassical}). \Cref{eq:EnergyError} predicts that
that the magnitude of this error is $2.8\%$ at $a_0 = 10$ and $0.3\%$
at $a_0 = 30$. This is comparable to the discrepancy shown in
\cref{fig:Ab-b} but for the largest $a_0$, where recoil corrections
to $N_{\gamma,\mcsp}$ take effect.

\Cref{eq:MostProbableNumber} indicates that the larger the lightfront
momentum transfer $f$, the more photons are absorbed from the external
field. Both the lightfront momentum transferred to an individual
photon and the number of emissions increase with $a_0$, so we expect
the absorption to increase as well. Integrating \cref{eq:MostProbableNumber}
weighted by the emission rate \cref{eq:LCFARate}, over all $f$ we find
that $N_L \sim a_0^2 \mathbb{I}^+$.
Here $\mathbb{I}^+$ is the total radiated lightfront momentum given by
\cref{eq:TotalLightfront}, which scales like $a_0^2$ in the classical limit.
Then we expect $N_L \sim a_0^4$, which agrees reasonably well with a
power-law fit to the data shown in \cref{fig:Ab-a}; we find $N_L
\propto a_0^{3.7}$ for both the QED and semiclassical results.
The true scaling is weaker than $a_0^4$ because of quantum corrections
that reduce the radiated power~\cite{Erber}.

We showed in \cref{sec:Lightfront} that the semiclassical method
predicts the large-$f$ part of the emission spectrum accurately even
for $a_0 = 5$. As this part of the spectrum is associated with the
the largest $\ell$, the agreement between the QED and semiclassical
results should be best for $\ell \gg 1$.
Four examples of the spectrum of probable $\ell$ are shown in
\cref{fig:Ab-c}. The agreement between the two is excellent for
$\ell > 10$, but the semiclassical method fails to capture the
small-$\ell$ part of the spectrum accurately. This is because
it localizes emission, thereby neglecting interference effects;
these suppress the emission probability for small $\ell$ and give
rise to the harmonic structure that can be seen in the QED spectra.

\subsection{Exemplary case}
\label{sec:Example}

	\begin{figure*}
	\subfloat[]{\label{fig:2d-a}\includegraphics[width=\linewidth]{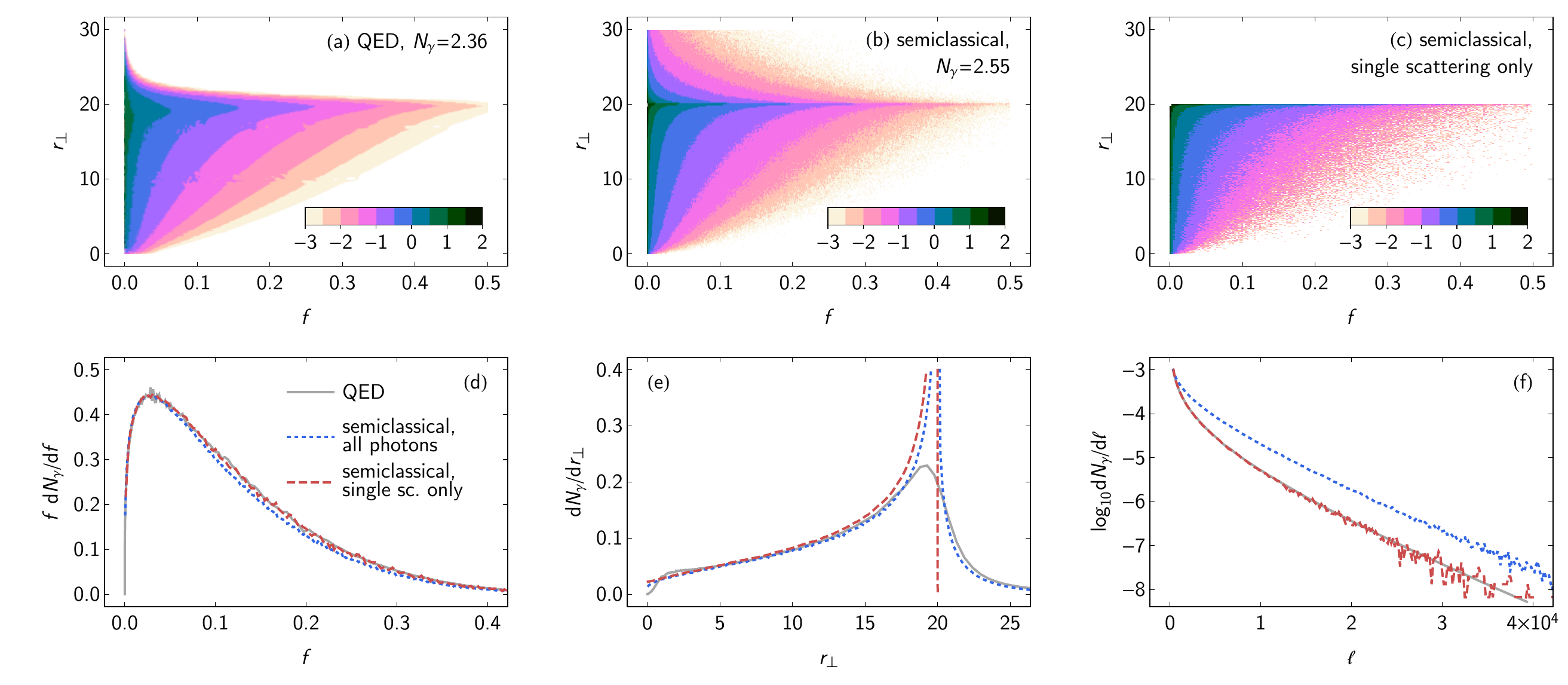}}
	\subfloat[]{\label{fig:2d-b}}
	\subfloat[]{\label{fig:2d-c}}
	\subfloat[]{\label{fig:2d-d}}
	\subfloat[]{\label{fig:2d-e}}
	\subfloat[]{\label{fig:2d-f}}
	\caption[Comparison]
		{(upper row) Double differential photon spectra
		$\rmd^2 N_\gamma / (\rmd f \rmd r_\perp)$ ($\log_{10}$-scaled)
		for a collision between an electron with $\gamma_0 = 1000$
		and a laser pulse with $a_0 = 20$ and $\tau = 2$: a)
		from QED, b) from semiclassical simulation, including all
		emissions, and c) from semiclassical simulation, filtered to
		single scattering only.
		(lower row) Single differential spectra for d) the lightfront
		momentum transfer fraction $f$, e) the scaled perpendicular
		momentum $r_\perp$ and f) the number of laser photons absorbed:
		results from QED (grey, solid) and from semiclassical simulation
		including all emissions (blue, dotted) and single scattering
		only (red, dashed).}
	\label{fig:2d}
	\end{figure*}

Finally, we present a comparison for a specific set of collision
parameters, drawing on the systematic results we have so far, to discuss
the role of multiple emissions. \Cref{fig:2d} shows the full set of
double- and single-differential photon spectra for lightfront momentum,
perpendicular momentum and absorption for a collision between an electron
with $\gamma_0 = 1000$ and a laser pulse with $a_0 = 20$ and $\tau = 2$.
The average number of photons is $N_\gamma = 2.36$ for these parameters
and therefore multiphoton effects should be taken into account.
However, as the QED calculation is performed only for single scattering,
we filter the semiclassical collision data to ensure the same physical
scenario is being compared. Now we can show the effect of this filtering
on the semiclassical results.

The double differential spectrum obtained when all emissions are taken
into account is shown in \cref{fig:2d-b}; when only single scattering
events are binned we obtain the spectrum shown in \cref{fig:2d-c}.
As discussed in \cref{sec:Perpendicular}, in the latter case we find
a sharp cutoff at $r_\perp = a_0$ as this is the largest angle between
the electron momentum and laser wavevector and we assume photons are
emitted in the collinear direction. Then photons with $r_\perp > a_0$
can only come from secondary scattering. Notice that while the QED
result (\cref{fig:2d-a}) is generally broader in the vertical direction,
the probability that $r_\perp > a_0$ diminishes with increasing $f$,
apparently justifying the assumption of collinear emission for $f \sim 1$.
The QED result is also smoother as it is free from the numerical
noise inevitable in Monte Carlo simulations. The single scattering
spectrum appears noisier as it represents only 20\% of the collision data.

\Cref{fig:2d-d} shows that the effect of the filtering on the lightfront
intensity spectrum is rather small. Nevertheless the agreement is
better when only single scattering is included. The spectrum in this
case is slightly harder, as secondary photons tend to be emitted with
smaller $f$. Taken in isolation, \cref{fig:2d-e} appears to suggest
that the agreement between the QED and semiclassical spectra is better
if we do not filter the collision data. The shape of the peak, if not
its maximum value, is actually captured better when all emissions are
included. This is coincidental. Recalling that we have assumed
collinear emission in the semiclassical approach,
we would expect the QED result for double Compton scattering to be
even broader in $r_\perp$.
In fact, the most significant difference is found when
we compare the probability that $\ell$ photons are absorbed from the
laser pulse in \cref{fig:2d-f}. The two semiclassical spectra have
the same integral (by construction, see \cref{sec:ComparisonBasis}),
therefore small values of $\ell$ must be suppressed when multiple
emissions are included. The probability that $\ell = 2\times10^4$,
for example, is $6\times$ larger when all emissions are accounted for.
Without the selection procedure we have introduced, it would not be
possible to compare these against the QED result in a consistent way.

\section{Discussion}

It is generally assumed that the semiclassical approach used in
modelling high-intensity laser interactions is valid when $a_0 \gg 1$
and $a_0^3/\chi_e \gg 1$. The precise value of $a_0$ for which these
conditions are satisfied depends, however, on the particular quantity
that is being calculated. We have shown that for $a_0$ as low as $5$,
semiclassical codes accurately capture the part of the emission
spectrum for which $f \sim 1$. On the other hand, there is still
a significant discrepancy in the total number of photons even when
$a_0 = 30$. The error is concentrated in the low-$f$ part of the
spectrum, i.e. photons with low energy and large angle; clearly
a semiclassical code should not be used to predict the result of
an experimental measurement in the spectral region $f \lesssim 2\chi_e/a_0^3$.
Improving these codes could be accomplished by calculating this
threshold and replacing the LCFA rate for photons with smaller
$f$~\cite{DiPiazza2017}, although this does require that
the external field can be treated as a slowly varying plane
electromagnetic wave.

Alternatively, it might be possible to bypass this problem by using
a PIC code, in which a hybrid description of the electromagnetic field
is employed. The Nyquist frequency associated with the finite spacing
of the grid naturally separates radiation into two components:
lower frequencies are resolved on the grid, i.e. classically,
and higher frequencies are treated as `photons', just as we outlined
in \cref{sec:Semiclassical}. This is why many implementations of
QED processes in PIC codes include the option of a low-frequency
cutoff below which macrophotons are not emitted~\cite{Epoch}.
It would be interesting to compare the QED results in this work
with the predictions of a semiclassical code in which the
radiation spectrum below a certain cutoff is obtained by Fourier
analysis of the Li{\'e}nard-Wiechart potentials in the far field.
This would ensure that the formation length is resolved at low $f$,
thereby capturing interference effects.
It is reasonable to expect a classical description to be appropriate
because both quantum corrections and the electron recoil should be
negligible for photons with $f \ll 1$.

While it is important to make these improvements at low $f$,
this part of the spectrum contributes negligibly to the
momentum change of the electron, which is dominated by photons with
large $f$. As the agreement between the QED and semiclassical spectra is
much better here, it is not surprising that we find
the average lightfront momentum loss predicted semiclassically to be
within a few per cent of the QED value even at $a_0 = 5$.
At $a_0 = 10$, for example, the error in the total number
of photons is 16\% for both $\tau = 2$ and 3, whereas in the
total radiated momentum $\mathbb{I}^+$ it is 1.5\%, an order
of magnitude better.
For the experimental parameters of \citet{Cole2018}
($a_0 \simeq 10$, $\chi_e \simeq 0.1$ and $\gamma_0 \simeq 1000$),
the lower limit on $f$ is equivalent to $\omega \lesssim 100$~keV;
this part of spectrum represents approximately 16\% of the total
number of photons but only $0.04\%$ of the total radiated energy,
using their parametrization of the spectrum and the measured
critical energy of 30~MeV.

This is encouraging for semiclassical or PIC-based modelling of
radiation reaction of an electron population.
In laser-beam interactions, the particle number density is generally low
enough that the radiation spectrum can be obtained by
incoherent summation over all photons emitted by the individual
particles; this allows the treatment of a beam with a spectrum
of energies and non-zero divergence. Although these two effects
will wash out, for example, detailed harmonic structure in
the momentum spectrum~\cite{Heinzl2010}, the overestimate at low $f$ will survive.
Nevertheless, as radiation reaction is an intrinsically
multiphoton process~\cite{DiPiazza2010}, proper benchmarking
requires the calculation of the higher-order diagrams shown
in \cref{fig:Feynman}. The selection and scaling scheme we have
presented here could easily be extended to comparisons with QED
calculations of double, triple etc. nonlinear Compton scattering.

Perhaps more important for the case of multiple emissions are the
comparisons we present for the perpendicular momentum spectra.
These test the assumption of collinear emission, which is distinct
from the LCFA. While the peak at $r_\perp \simeq a_0$ is common
to both QED and semiclassical results, the width of the distribution
around this point is not captured semiclassically. The angle
at which the photon and electron travel after the scattering
affects their quantum parameter and so the rates at which secondary
processes occur. Even if the change to the rates is small, the
cumulative error could become large if the multiplicity is high.
Accurate modelling of the angular spectrum is important because,
for example, the transverse broadening of an electron beam has been
proposed as a signature of quantum effects on radiation reaction~\cite{Green}.
This broadening would be in addition to that from the
finite beaming of the radiation, and any initial divergence
of the beam (a few milliradians for the laser-wakefield-accelerated
electron beams reported by~\cite{Cole2018,Poder}).
It would also be important for the study of QED avalanches, in which
even a single electron accelerated by counterpropagating lasers
can seed the creation of a critically dense electron-positron pair
plasma.
One possible approach would be to implement an angularly-resolved
LCFA rate that includes the finite $1/\gamma$ beaming of the scattered
photon (see \cite{BKS} for example).

We have also found that the absorption of energy from the background
is dominated by the `classical' component, i.e. $\vec{j}\cdot\vec{E}$
work done by the external field in accelerating the scattered
electron. This is assumed to be the case in PIC modelling of
QED avalanches and suggests that a classical treatment of backreaction
is reasonably accurate at high intensity. (Recall that in these codes
the fields and currents are evolved self-consistently, but classically.)
There remains the question of the `missing' absorption we discussed in
\cref{sec:Absorption}. On the one hand the fraction of the total
depletion this represents diminishes with increasing $a_0$; however,
if this error does arise on a `per-emission' basis, the increased
multiplicity at high $a_0$ could mean that it becomes significant.
It is reasonable to expect a causal relationship between the absorption
discrepancy and the assumption of collinear emission. This is something
we will consider in future work.

\section{Conclusions}

In this work we have presented benchmarking of semiclassical simulations
against exact QED results for nonlinear Compton scattering in an
intense laser pulse, using a method that guarantees that we compare precisely
the same physical scenario.

The differential spectra agree both qualitatively and quantitatively
in the dynamically important region $f \gtrsim 2\chi_e/a_0^3$
that dominates the electron recoil and absorption from the laser fields.
We find that the lightfront momentum loss and number of absorbed photons
from semiclassical simulations are within a few percent of the exact QED
results for $a_0 > 5$.
However, improvements are clearly called for at low $f$, where the
LCFA breaks down, and in the angular distribution, where the agreement
is only qualitative due to the assumption of collinear emission in
the simulations.

It remains to be seen whether improving these will lead to significant
differences in the results of simulations of laser-plasma interactions.
In deciding what is most important we should be guided by further
comparison with QED calculations that include multiple emissions.
These will place more stringent limits on the validity of the
approximations that underpin the semiclassical approach.
The fact that experimental exploration of the strong-field, multiphoton
regime ($a_0 \sim 10$, $\chi_e \sim 0.1$, $N_\gamma \sim 10$) is
already underway makes this an urgent question.

\begin{acknowledgments}
The authors thank T. Heinzl and A. Ilderton for useful discussions
and acknowledge support from
the Knut and Alice Wallenberg Foundation (T.G.B., M.M.),
the Science and Technology Facilities Council, grant ST/G008248/1 (D.S.),
the US DOE under Contract No. DE-AC02-05CH11231 (S.S.B.),
and the Swedish Research Council, grants 2013-4248 and 2016-03329 (M.M.).
\end{acknowledgments}

\bibliography{references}

\end{document}